\documentclass[12pt]{article}
\RequirePackage[OT1]{fontenc}
\RequirePackage{amsthm,amsmath}
\usepackage{accents}
\usepackage{breqn}
\usepackage{soul}
\usepackage[numbers,sort&compress]{natbib}
\RequirePackage[citecolor=blue,urlcolor=blue]{hyperref}
\usepackage{hyperref}
\usepackage{amsfonts}
\usepackage{amssymb}
\usepackage{graphics}
\usepackage{setspace}
\usepackage{physics}

\newtheorem{definition}{Definition}[section]

\numberwithin{equation}{section}
\newcommand{\bsigma}{{\boldsymbol \sigma}}

\def\b1{{\mathbf 1}}
\def\b0{{\mathbf 0}}

\def\bA{{\boldsymbol A}}

\def\bn{{\boldsymbol n}}

\def\bv{{\boldsymbol v}}
\def\bV{{\boldsymbol V}}
\def\br{{\boldsymbol r}}

\def\bH{{\boldsymbol H}}

\def\bU{{\boldsymbol U}}

\def\bbR{{\mathbb R}}

\def\bbI{{\mathbb I}}

\begin{document}

%\begin{frontmatter}
\title{Qubit thermalization by random pulses: Asymptotic state factorization}
\author{Henryk Gzyl\\
Centro de Finanzas IESA, Caracas, Venezuela.\\
 henryk.gzyl@iesa.edu.ve}
\date{}
 \maketitle

\setlength{\textwidth}{4in}
%\end{frontmatter}

\vskip 1 truecm
\baselineskip=1.5 \baselineskip \setlength{\textwidth}{6in}

\begin{abstract}
Here we consider an analytically tractable model of a two level quantum system subject to random shocks and prove that it decays asymptotically to a trivial state, that is, to a state in which the two levels have equal probability of occupation. In a two qubit system, if the shocks affect each qubit independently, the equilibrium density matrix becomes a simple product of the one qubit equilibrium density matrices regardless of the nature of the initial state. This has potential applications to entangles qubits in quantum computers.
\end{abstract}


\begin{spacing}{0.5}
\small
\tableofcontents
\end{spacing}

\section{Introduction and Preliminaries} 
If we type ``disentanglement'' in our favorite browser, we will be provided with links to almost any type of human endeavor. If you refine your search to ``disentanglement in quantum mechanics'', you will find references to works in very many domains of application of quantum mechanics. It thus makes sense to recall the meaning of separability, correlation and entanglement.

When dealing with a system composed of two subsystems, in the Schr\"{o}dinger description one says that the state $\Psi$ of the whole system is an entangled state when there do not exit two states $\Psi_1$ and $\Psi_2$ such that $\Psi=\Psi_1\otimes\Psi_2.$ In the Heisenberg representation things are more complicated. From \cite{W} or \cite{Horod}, we recall that:

\begin{definition}\label{separ0}
Consider a composite quantum system made up $K$ subsystems. A quantum state of the composite system, described by a density matrix $\rho$ is \textbf{separable} if it can be written as:
\begin{equation}\label{separ1}
 \rho = \sum_{k=1}^K\mu_k|\psi_{k_1}\rangle\langle\psi_{k_1}|\otimes....\otimes|\psi_{k_1}\rangle\langle\psi_{k_1}|.\
\end{equation}
The definition in terms of density matrices is similar, but more general because not all densities represent pure states. The state $\rho$ is separable, classically correlated, or mixed, whenever:
\begin{equation}\label{separ2}
\rho = \sum_{k=1}^K\mu_k\,\rho_{k_1}\otimes...\otimes\rho_{k_N}.
\end{equation}
Here the $\{\mu_k:k=1,...,K\}$ are positive numbers such that $\sum\mu_k=1.$  Similarly, the vectors $\{|psi_{k_1}\rangle: k=1,...,K\}$ are vectors in the Hilbert spaces in the Schr\"{o}dinger representation of the states of each subsystem. Similarly, 
$\{\rho_k: k=1,...,K\}$ are states in the corresponding Heisenberg representation.
A state $\rho$ is called \textbf{entangled} if it is not separable. 
\end{definition}

Also, making use of the Bloch representation (or using the Pauli spin matrices as basis vectors for the class of density matrices), we shall see that the asymptotic behavior of any density matrix can be obtained from the asymptotic behavior of action of the time evolution operator acting on the Pauli matrices. To prove entanglement may not be easy. Nevertheless, there is a partial test, the \textit{purity} test, that discriminates between pure and non-pure states. Given a state $\rho,$ it is a pure state if $\rho^2=\rho.$ Therefore, if $tr(\rho^2)<tr(\rho)=1,$ the state cannot be a pure state. The example presented in \eqref{ini1}-\eqref{ini1}-\eqref{ini1} below is an example of a separable, but non-pure state.

The reason for considering density matrices is clear. Even though we work out the time evolution of the system in the Schr\"{o}dinger representation, when we need to make predictions about expected values, we need the density matrix. That is, first we need to compute a trace with respect to the density matrix for each possible random evolution, and then average over all possible random evolutions. 
 
The aim of this paper is to solve the time evolution of qubit subject to a random perturbation whose effect is to flip the unperturbed basis states, and the switching occur at exponentially distributed time intervals. Once we finde the explicit time evolution for any random sequence of shocks and any initial state in the Schr\"{o}dinger representation, we transport it to the Heisenberg representation to determine the corresponding time evolution of  density matrices.

After that, we consider two identical, non-interacting, two level sub-systems (qubits, say), in an arbitrary initial state (density matrix), immersed in the heat bath described above. We show that, in that model of heat bath, the system evolves into a trivial product state.

For applications in quantum information theory and quantum computer design, and worrisome issue is the presence of thermal nose destroying the entanglement necessary for the quantum codes, therefore, a simple model displaying this phenomenon might be of interest 

\subsubsection{The basic assumptions of the model}
We consider two identical non-interacting two level systems, coupled through their initial density matrix. The time evolution of each system is described by the same Hamiltonian $\bH,$ and the effect of the shocks on the states are represented by the following matrices on the canonical basis.

\begin{equation}\label{eq1}
\bH = \begin{pmatrix}
       \epsilon_1 & 0 \\
				0   & \epsilon_2\end{pmatrix}\;\;\;\;\bV = \begin{pmatrix}
                                                 0 & 1 \\
			                                        	 1 & 0\end{pmatrix} .   
\end{equation}

In \eqref{eq1} $\bH$ represents the Hamiltonian of each component, and $\bV$ the unitary operator describing the effect of the shocks produced by the heat bath. The basis vectors for each state	will be denoted by the customary $|1\rangle$ and $|2\rangle,$ and such that $\bH|i\rangle=\epsilon_i|i\rangle$ for $i=1,2.$ And the effect of $\bV$ on the same basis vectors is given by $\bV|1\rangle=|2\rangle$ and $\bV|2\rangle=|1\rangle.$ 

The model of the heat bath considered here goes back to Clauser and Blume in \cite{CB}. The system starts at some initial state $\Psi$ (or $|\Psi\rangle$ depending of notational convenience), evolves according to $\bH$ during a random time $\tau_1,$ at which the perturbation given by $\bV$ acts instantaneously and flips the spin of the system. After the shock, the system again evolves according to $\bH$ during another random time $\tau_2$ at the end of which another instantaneous state flipping according to $\bV$ occurs. This is repeated infinitely many times. The instantaneous perturbation may be a more elaborate operator changing the state of the system instantaneously, but the model we consider here allows for full analytical computations and suffices to convey the essence of the thermalization process.

The underlying proposal in \cite{CB} can be extended to a theory of random quantum evolutions of systems with a finite number of levels as in \cite{Gz}.

The assumptions underlying the model are the following.
\begin{itemize}
\item The two sub-systems do not interact. That is, each of them evolves according to the Hamiltonian in \eqref{eq1}.
\item The heat bath affects each of them independently of the other. That is,the occurrence of the shock that each suffers are independent of each other (we may assume that they are equally distributed). 
\end{itemize}
The second assumption is natural if we think of the subsystems as spatially separated. Thus even if they respond similarly to the heat bath, their responses are independent of each other

\subsubsection{Statistical properties of the noise}
We suppose that the times between shocks are independent and exponentially distributed, with frequency $\lambda$ (or mean inter arrival time $1/lambda.$  If we write $T_n=\sum_{k}\tau_k,$ to denote the time of occurrence of the $n$-th shock, and define the shock counting process $N(t)$ by $\{N(t)=n\}\;\Leftrightarrow \;\{T_n\leq t\leq T_{n+1}\}.$ One stes $T_0=0$ to mean that when the experiment is started, no shocks have occurred.

Statistically speaking, the process is characterized by saying that $N(0)=0,$ that for any $t,s\geq 0,$ the random variables $N(t+s)-N(s)$ has the same probability distribution as $N(t),$ and that for any $t_1<t_2<t_3<t_4,$  the increments $N(t_4)-N(t_3)$ and  $N(t_2)-N(t_1)$ are statistically independent. Also, that for any $t\geq 0,$
\begin{equation}\label{PD1}
P(N(t)=n) = \frac{(\lambda t)^n}{n!}e^{t\lambda}.
\end{equation}
Not only that, if $0<T_1<T_2<...<T_n<;;;$ denote the (random) jump times of $N(t),$ then the times $\tau_1=T_1$ and $\tau_n=T_n-T_{n-1}$ are independent and exponentially distributed with parameter $\lambda.$ Besides, $\{N(t)=0\} =\{t<T_1\}$ and $\{N(t)=n\}=\{T_n\leq t<T_{n+1}\}.$ To finish, the conditional distribution of $T_1,T_2,...,T_n$ given that $\{N(t)=n\}$ is uniform in the sector $0,t_1<t_2<...<t_n<$ of the positive orthant in $\bbR^n$ with density $n!/t^n.$ This means that, for any, continuous say, function $h(T_1,T_2,...,T_n)$ we have

\begin{equation}\label{PD2}
E[h(T_1,T_2,...,T_n)|N(t)=n] = \frac{n!}{t^n}\int_0^t dt_n\int_0^{t_n}dt_{n-1}\int_0^{t_n-2}...\int_0^{t_2}dt_H(t_1,t_2,...,t_n),
\end{equation}
We shall put this to a good use below.

The stochastic nature of this thermal bath model is simple but intuitive, which together with the spin flipping perturbation, makes the model analyticall tractable.  Many properties and applications of the Poisson process are discussed in \cite{F} or \cite{Pin}, and for ma more physical flavor \cite{vK} or \cite{L}. For our needs, it suffices to say that the process $N(t)$ has right continuous trajectories with left limits, that it is constant between jumps, and that all jumps are of size $1.$ 

\subsection{A quick review of the literature}
The literature is about disentanglement is rather large. To put it into context, for a couple of reviews about entanglement, consider \cite{TWD} and \cite{Horod}. The overlapping themes in the list are relationship to decoherence, to blind source separation,  Bell's inequalities,  the nature of the two-state system, the nature of the thermal bath. To begin with, consider \cite{SMDZ} for the underlying issue of entanglement, and \cite{AWL}, \cite{KK} for the creation and stability of entangled sates. For the possible existence of disentanglement free states see \cite{YG}. For different theoretical issues related to disentangling consider \cite{LZY}, \cite{LSKR}, \cite{Daj}, \cite{ACXZ}, \cite{GBRSK}, \cite{BH}, \cite{ZDF}. For the dynamics of disentangling see \cite{SMDZ} and \cite{WHBH}. For disentanglement after spontaneous emission \cite{San}, and for different models of the thermal bath see \cite{IKZ}, \cite{CZ}, \cite{QTII}, \cite{MW}, and \cite{MHO}. For a review of both theoretical and experimeantal results, and several models of thermal bath see \cite{AMD} and for thermalization of a photonic cube see \cite{Oetal} and \cite{CCS} for the thermalization of two level quantum systems.

For the relationships between disentanglement decoherence  and dephasing, consider \cite{HO}, \cite{YE}, \cite{AJ}, \cite{ZD}, \cite{WNR}, \cite{DH}. For disentanglement and teleportation see \cite{GKRSS}, \cite{YY}. For a relationship to Bell's inequalities \cite{LL} and for a potential application to blind source separation \cite{DD}.

\section{The Blume-Clauser model for the time evolution of the density matrix}
The starting point is the time evolution operator in the Schr\"{o}dinger picture. Before a random shock happens, the time evolution operator is $U_0(t)=\exp(-it\bH/h)$ describes the time evolution of the unperturbed system. We use $U(t)$ to denote the time evolution operator of the perturbed system. Suppose that a random time $T_1$ a random shock occurs, since before the first shock the evolution was according to $U_0,$ cecause of continuity we have:
\begin{equation}\label{TE1}
U(T_1-) = lim_{t\,\uparrow\, T_1}U(t) = lim_{t\,\uparrow\, T_1}U_0(t)=U_0(T_1).
\end{equation}
Or if $\Psi(0)$ denotes the initial state, right before the collision the state is
\begin{equation}\label{TE1a}
\Psi(T_1-) = U_0(T_1-)\Psi(0).
\end{equation}
Right after the (instantaneous) shock at $T_1$ the time evolution operator of the random system is
\begin{equation}\label{TE2}
U(T_1) = \bV U(T_1-) = \bV U_0(T_1).
\end{equation}
And in terms of the state vector:
\begin{equation}\label{TE2a}
\Psi(T_1) = \bV\Psi(T_1-) = \bV U_0(T_1-)\Psi(0).
\end{equation}
These equations capture the essence of mathematical description of the discontinuity of the state vector and that of the evolution operator. To continue, during the interval $T_1\leq t < T_2,$ the system again evolves according to $U_0,$ that is
\begin{equation}\label{TE3}
U(t) = U_0(t-T_1)U(T_1) = U_0(t-T_1)\bV U_0(T_1).
\end{equation}
When the next shock occurs at $T_2:$
$$ U(T_2-) = lim_{t\,\uparrow\, T_2}U(t)=lim_{t\,\uparrow\, T_2}U_0(t-T_1)\bV U_0(T_1)=U_0(T_2-T_1)\bV U_0(T_1)$$
Right after the shock $U(T_2) = \bV U(T_2-) = \bV U_0(T_2-T_1)\bV U_0(T_1).$ 
And so on and on.\\
 These comments can be summarized as
\begin{equation}\label{TE4}
U(t) = \left\{\begin{array}{ccc}
             Evolution\; Operator  & Number\; of\; Shocks & Probability\\
               U_0(t)      &  \{N(t)=0\}      & e^{-\lambda t}\\
           U_0(t-T_n)\bV U_0(T_n-T_{n-1}\bv...\bV U_0(T_1) & \{N(t)=n\} & e^{-\lambda t}\frac{(\lambda t)^n}{n!}.
					\end{array}\right.
\end{equation}
If we introduce the notations $I_{\{N(t)=n\}}$ to denote the binary, random variable that assumes value $1$ when the event $\{N(t)=n\}$ happens and $0$ if not, the previous definition can be written in a more convenient form:

\begin{equation}\label{TE5}
U(t) = \sum_{n=0}^\infty U_0(t-T_n)\bV U_0(T_n-T_{n-1})\bV...\bV U_0(T_1)
I_{\{N(t)=n\}}.
\end{equation}
Notice that the first term in the sum, corresponding to $n=0$ is simply 
$U_0(t)I_{\{N(t)=0\}}.$ When considering states (density matrices) in the Heisenberg representation, it will prove convenient to introduce the notations

\begin{equation}\label{not2.1}
\bV^{\sharp}(\bA) = \bV\bA\bV^{\dagger}\;\;\;\;\bU_0^{\sharp}(\bA) = \bU_0\rho\bU_0^{\dagger}\;\;\;\bH^\times(\bA)=[\bH,bA]
\end{equation} 
where $\bA$ stands for any $2\times2$ matrix. Omitting the argument of the operators, and keeping in mind that $I_{\{N(t)=n\}}I_{\{N(t)=m\}}=0$ when $n\not=m,$ the time evolution in the Heisenberg representation is given by
	
\begin{equation}\label{TE6}
U^{\sharp}(t) = \sum_{n=0}^\infty U^{\sharp}_0(t-T_n)\bV^{\sharp} U_0^{\sharp}(T_n-T_{n-1})\bV^{\sharp}...\bV^{\sharp} U_0^{\sharp}(T_1)I_{\{N(t)=n\}}.
\end{equation}
For each state of the heat bath, that is, for each random sequence of shocks, for each intial state $\rho(0),$ the (random) expected value of an observable $\bA$ is given by:
\begin{equation}\label{EV1}
\tr\big(\rho(t)\bA\big) = \tr\big(\bA U^{\sharp}(t)(\rho(0))\big) = \tr\big(\bA U(t)\rho(0)U(t)^{\dagger}\big).
\end{equation}					
If we denote by $\langle\rho(t)\rangle_{av}$ the average value of $\rho(t)$ over all states of the heath bath, then the expected value of any observable is obtained by averaging \eqref{EV1} over all states of the heath bath (that is over all possible random sequences of shock times) is:	 		

\begin{equation}\label{EV2}
\langle\tr\big(\rho(t)\bA\big)\rangle_{ave} = \tr\big(\bA\langle U^{\sharp}(t)\rangle_{av}(\rho(0)\big).
\end{equation}

 Let us put $u(t)=\langle U^{\sharp}(t)\rangle_{ave}$ and use \eqref{TE6} and \eqref{PD2} to write $u(t)$ as

\begin{equation}\label{TE7}
u(t) = e^{-t\lambda}\sum_{n=0}^\infty \lambda^n \int_0^t dt_n U_0^{\sharp}(t-t_n)\bV^{\sharp}\int_0^{t_n} dt_{n-1}U_0^{\sharp}(t_n-t_{n-1})\bV^{\sharp}\int_0^{t_n-2}...\bV^{\sharp}\int_0^{t_2}U_0^{\sharp}(t_1)dt_1.
\end{equation}
In line with the comment right after \eqref{TE5}, it is convenient to keep in mind that the first term in the summation in \eqref{TE7} is 
$U_0^{\sharp}(t)\exp(-t\lambda).$ Now apply $i\hbar\partial/\partial t$ to both sides of \eqref{TE7} to obtain
$$
\begin{aligned}
&i\hbar\frac{\partial u(t)}{\partial t} = -i\hbar\lambda u(t) + \bH^{\times}U_0^{\sharp}(t)e^{-t\lambda}  \\
       &e^{-t\lambda}i\hbar\frac{\partial}{\partial t}\sum_{n=1}^\infty \lambda^n \int_0^t dt_n U_0^{\sharp}(t-t_n)\bV^{\sharp}\int_0^{t_n} dt_{n-1}U_0^{\sharp}(t_n-t_{n-1})\bV^{\sharp}\int_0^{t_n-2}...\bV^{\sharp}\int_0^{t_2}U_0^{\sharp}(t_1)dt_1.
\end{aligned}
$$
The derivative of each term of the summation involves two terms. When differentiating with respectt to the upper limit of integration ve obtain
$$\bV^{\sharp}\int_0^{t} dt_{n-1}U_0^{\sharp}(t_n-t_{n-1})\bV^{\sharp}\int_0^{t_n-2}...\bV^{\sharp}\int_0^{t_2}U_0^{\sharp}(t_1)dt_1,$$
which corresponds to the previous term in the summation. Inserting back in the summation and relabeling the summands we obtain
$$\lambda\bV^{\sharp}u(t).$$ 
When differentiating the $U_0^{\sharp}(t-t_n)$ under the integral sign we obtain $-i\bH^{\times}/\hbar$ applied to the same term. 
\[
\begin{aligned}
&\frac{-i}{\hbar}\bH^{\times} \sum_{n=1}^\infty \lambda^n \int_0^t dt_n U_0^{\sharp}(t-t_n)\bV^{\sharp}\int_0^{t_n} dt_{n-1}U_0^{\sharp}(t_n-t_{n-1})\bV^{\sharp}\int_0^{t_n-2}...\bV^{\sharp}\int_0^{t_2}U_0^{\sharp}(t_1)dt_1 \\
&= \frac{-i}{\hbar}\bH^{\times} \big(u(t)e^{t\lambda}- U_0^{\sharp}(t)\big).
\end{aligned}
\]
Putting all of this back together, we obtain the same equation as in the example in \cite{Gz}, except that now all the steps are shown explicitly, that is
\begin{equation}\label{TE8}
i\hbar\frac{\partial u}{\partial t} = \bH^{\times}u + i\hbar\lambda\big(\bV^{\sharp}-\bbI)u.
\end{equation}

\subsection{The integration of \eqref{TE8}}
For not to use that many subscripts, let us write $\rho(t)=u(t)(\rho(0))$ as:
$$ \rho(t) = \begin{pmatrix} a & b\\c & d\end{pmatrix}.$$
This us a hermitian matrix of trace $1.$ A simple computation shows that

\begin{equation}\label{TE9}
\bH^\times\rho(t) = (\epsilon_1-\epsilon_2)\begin{pmatrix} 0 & b\\-c&0\end{pmatrix},\;\;\;\;\;\;\;\big(\bV^{\sharp}-\bbI\big)\rho = \begin{pmatrix} d-a & c-b\\b-c & a-d\end{pmatrix}.
\end{equation}

Equating terms on both sides of \eqref{TE5} we obtain the two systems:
\begin{equation}\label{TE10}
\begin{aligned}
\frac{\partial a}{\partial t} &= \lambda(d-a)\\
\frac{\partial d}{\partial t} &= -\lambda(d-a) .
\end{aligned}
\end{equation}
In the system for $c$ and $b,$ we put $\omega=(\epsilon_1-\epsilon_2)/\hbar.$ As above, we obtain
\begin{equation}\label{TE11}
\begin{aligned}
\frac{\partial b}{\partial t} &= -i\omega b - \lambda(b-c)\\
\frac{\partial c}{\partial t} &= i\omega c + \lambda(b-c).
\end{aligned}
\end{equation}
The perturbation matrix $\bV$ is such that the dynamics diagonal and the non-diagonal elements is separated. That system is easy to integrate. Note that adding and subtracting the two equations we obtain

\begin{equation}\label{TE12}
\begin{aligned}
&\frac{\partial (a+d)}{\partial t} = 0 \;\;\Longrightarrow \;\;\;a(t)+d(t)=a(0)+d(0) .\\
&\frac{\partial (a-d)}{\partial t} =-2\lambda(a-d)\;\;\;\Longrightarrow \;\;\;a(t)-d(t)=(a(0)-d(0))e^{-2t\lambda}.\
\end{aligned}
\end{equation}

It is not hard to verify that adding and subtracting and differentiating once more the system \eqref{TE11},  we obtain the system:

\begin{equation}\label{TE13}
\begin{aligned}
&\frac{\partial^2 (b+c)}{\partial t^2}+2\lambda\frac{\partial(b+c)}{\partial t}+\omega^2(b+c) = 0.\\
&\frac{\partial^2 (b-c)}{\partial t^2}+2\lambda\frac{\partial(b-c)}{\partial t}+\omega^2(b-c) = 0.
\end{aligned}
\end{equation}
 The initial condition for the derivatives is obtained evaluating \eqref{TE12} at $t=0:$ 
\begin{equation}\label{TE13a}
\begin{aligned}
&\frac{\partial (b+c)}{\partial t}(0)=-i\omega(b- c)(0). \\
&\frac{\partial (b-c)}{\partial t}(0)=-i\omega(b+c)(0)-2\lambda(b-c).
\end{aligned}
\end{equation}
 If we write $x(t)$ for the unknown in any of \eqref{TE13a}, then standard procedure leads to:
\begin{equation}\label{TE13b}
\begin{aligned}
&\;\;\;x(t) = \frac{\dot{x}(0)-r_{-}x(0)}{r_{+}-r_{-}}e^{r_{+}t} + \frac{-\dot{x}(0)+r_{+}x(0)}{r_{+}-r_{-}}e^{r_{-}t}.\\
\mbox{Where}&\;\;\;r_{\pm} =-\lambda\pm\sqrt{\lambda^2-\omega^2}\;\;\;\mbox{are the characteristic roots of the equations.}
\end{aligned}
\end{equation}
Now, undoing all tha changes of variables, the explicit time dependence of $a(t),b(t),c(t),d(t)$ turns out to be: 

\begin{equation}\label{TE13.1}
\begin{aligned}
&a(t) = \frac{a(0)}{2}\big( 1+e^{-2t\lambda}\big)+\frac{d(0)}{2}\big( 1-e^{-2t\lambda}\big).\\ 
&d(t) = \frac{a(0)}{2}\big( 1-e^{-2t\lambda}\big)+\frac{d(0)}{2}\big( 1 + e^{-2t\lambda}\big). \\
&b(t)=\frac{1}{r_{+}-r_{-}}\bigg(\big[\dot{b}(0)-r_{-}b(0)\big]e^{tr_{+}}+\big[-\dot{b}(0)+r_{+}b(0)\big]e^{tr_{-}}\bigg)\\
&c(t)=\frac{1}{r_{+}-r_{-}}\bigg(\big[\dot{c}(0)-r_{-}c(0)\big]e^{tr_{+}}+\big[-\dot{c}(0)+r_{+}c(0)\big]e^{tr_{-}}\bigg).
\end{aligned}
\end{equation}
We already mentioned that the initial conditions $\dot{b}(0),\dot{c}(0)$ are to be obtained from \eqref{TE11}. Notice that, whatever the sign of the discriminant $\lambda^2-\omega^3,$ the long time behavior of the exponential is decaying at rate $\lambda.$ Therefore, letting $t\to\infty$ we obtain that asymptotically the non-diagonal terms of $\rho(t)$ vanish away, and since $a(0)+d(0)=1$ if the trace of $\rho(0=1,$ therefore:

\begin{equation}\label{TE13.2}
\rho(t) = \begin{pmatrix} a(t) & b(t)\\c(t) & d(t)\end{pmatrix}\;\;\;\longrightarrow \;\;\;\begin{pmatrix} \frac{1}{2}(a(0)+d(0))  & 0 \\ 0 & \frac{1}{2}(a(0)+d(0))\end{pmatrix}.
\end{equation}

An interesting twist appears when noticing that in the passage from \eqref{TE8}-\eqref{TE9} to \eqref{TE13.1} no use was made of the nature of the initial matrix. Next, we integrate the system assuming the initial conditions correspond to the  Pauli matrices $\sigma_i: i=1,2,3.$  These are:

\begin{equation}\label{pauli}
\sigma_1 = \begin{pmatrix} 0 & 1\\ 1 & 0\end{pmatrix}\;\;\;\sigma_2 = \begin{pmatrix} 0 & -i\\ i & 0\end{pmatrix}\;\;\;\sigma_3 = \begin{pmatrix} 1 & 0\\ 0 & -1\end{pmatrix}.
\end{equation}
 We add $\sigma_0=\bbI,$ and with an obvious abuse of notation, denote by $\sigma_\mu(t)$ (with $\mu=1,1.2.3$), the result of adapting \eqref{TE13.1} to the case in which the initial condition $\rho(0)=\sigma_mu.$ After some tedious labor, we obtain:

Applying \eqref{TE13.1} successively to each Pauli matrix we obtain
\begin{gather}
\sigma_0(t) = \sigma_0(0) = \bbI.\;\;\;\label{pauli0}\\ 
\sigma_1(t)= \frac{1}{r_{+}-r_{-}}\bigg(\omega\big[e^{tr_{+}}-e^{tr_{-}}\big]\sigma_2-\big[r_{+}e^{tr_{-}}-r_{-}e^{tr_{+}}\big]\sigma_1\bigg).\;\;\;  \label{pauli1}\\
\sigma_2(t) = \frac{1}{r_{+}-r_{-}}\bigg(\omega\big[e^{tr_{-}}-e^{tr_{+}}\big]\sigma_1-\big[(r_{-}+2\lambda)e^{tr_{+}}-(r_{+}+2\lambda)e^{tr_{-}}\big]\sigma_2\bigg) \;\; \label{pauli2}\\
\sigma_3(t) = e^{-2\lambda t}\sigma_3.   \label{pauli3}
\end{gather}
The fact that the identity matrix is constant just means that the random evolution preserves the unitarity when the average is taken over all states of the environment. These identities may prove useful when studying the time evolution of a state like 
\eqref{separ1} if we translate from the bra-ket notation to the Bloch representation $\rho=(1/2)\big(\bbI+\br\cdot\bsigma\big)$ for pure states. 

Of course, the same goes for \eqref{separ2} if the initial state of the two qubit system is a classical correlation of non-pure states.

\section{Examples of asymptotic decorrelation}
We begin with a simple example of a separable but not pure state, that becomes asymptotically pure. After that we consider a general state and examine under what conditions it becomes asymptotically uncorrelated or perhaps pure. 

\subsection{A simple separable state}
The following state is separable (in the Heisenberg representation):
\begin{equation}\label{ini1}
\rho(0) =\frac{1}{2}\big(|1\rangle\langle 1|\otimes|2\rangle\langle 2| + |2\rangle\langle 2|\otimes|1\rangle\langle 1|\big),
\end{equation}
In matrix form, the state is given by:
\begin{equation}\label{ini2}
\rho(0) = \frac{1}{2}\left(\begin{pmatrix}
                            1  & 0\\
														0 & 0\end{pmatrix}\otimes\begin{pmatrix}
                            0  & 0\\
														0 & 1\end{pmatrix} + \begin{pmatrix}
                                                 0  & 0\\
														                     0 & 1 \end{pmatrix}\otimes\begin{pmatrix}
                                                                   1  & 0\\
														                                       0 & 0\end{pmatrix}\right).
\end{equation}
The left factor of the tensor product standing for a density matrix of the first sub-system and the right factor standing for a density matrix of the second sub-system. If we work out the tensor products, we obtain the $4\times4$ density matrix{

\begin{equation}\label{ini3}
\rho(0) = \begin{pmatrix}
           0 & 0 & 0 & 0\\
					 0 & 1/2 & 0 &\\
					 0 & 0 & 1/2 & 0\\
					 0 & 0 & 0 & 0\end{pmatrix}
\end{equation}											
This is a proper density matrix describing an separable, non-pure state of the composite system, and it is also invariant under the exchange of the sub-systems, as required for systems composed of identical sub-systems. Since $tr(\rho^2)=1/2,$ therefore this $\rho$ is not a pure state. To examine its time evolution under the action of the random environment, first notice that:
$$
\begin{aligned}
& |1\rangle\langle 1| \;\;\;\mbox{in the Bloch representations is}\;\;\;\frac{1}{2}(\bbI+\sigma_3).\\
& |2\rangle\langle 2| \;\;\;\mbox{in the Bloch representations is}\;\;\;\frac{1}{2}(\bbI-\sigma_3).
\end{aligned}
$$
Therefore, \eqref{ini1} can be rewritten as
\begin{equation}\label{ini1.1}
\rho(0) = \frac{1}{8}\bigg((\bbI+\sigma_3)\otimes(\bbI-\sigma_3)+(\bbI-\sigma_3)\otimes(\bbI+\sigma_3)\bigg).
\end{equation}
As the perturbation acts independently on each quibit, invoking \eqref{pauli3}  we obtain that
\begin{equation}\label{ini1.2}
\rho(t) = \frac{1}{8}\big(\bbI\otimes\bbI +\bbI\otimes\bbI\big) = \frac{1}{4}\bbI\otimes\bbI.
\end{equation}
In this simple example, both states of the two qubit system have the same probability of occurrence.

\subsection{The general case}
Here we work out the same example in two different ways that differ in the form used to express the initial state $\rho(0).$ The first one is based on the fact that $\sigma_0\equiv\bbI,$ and the Pauli matrices $\sigma_i,i=1,.2,3$ form a basis for the space of $2\times2$-Hermitian matrices, and the second is the representation of separable states introduces in \eqref{separ2}. Furthermore, we assume that the two particles (or subsystems) are identical, and the representation should be ivariant ubder the exchange of particles (or labels). The two representations are:

\begin{gather}
\rho(0)=\frac{1}{4}\sum_{\mu,\nu=0}^3C_{\mu,\nu}\sigma_{\mu}\otimes\sigma_{\nu}\;\;\; \label{paurep}\\
\rho(0)=\sum_{k=1}^Kp_k\rho_{1,k}\otimes\rho_{2,k} \;\;\;\label{seprep}
\end{gather}
The first representation stands for the expansion of an arbitrary density matrix in terms of the basis for the density matrices of the bipartite system. The second representation may be a separable state if the $\rho_{i,k}$ is a pure state, otherwise $\rho(0)$ is a classically correlated state. Next, we examine each the asymptotic state for each representation. Let us rewrite the first one as

The factor $1/4$ has been introduced to set $C_{0,0}=1$ henceforward. We used assumption of symmetry with respect to the exchange of particles to simplify the representation. If we now use the assumption that the environment acts independently on each particle. Therefore, each item in the tensor products evolves in time according to \eqref{pauli0}--\eqref{pauli3}, then the asymptotic form of the state \eqref{inipau1} is:

\begin{equation}\label{asypau1}
\rho(\infty)= \frac{1}{4}\big(\bbI\otimes\bbI\big) = \frac{1}{2}\bbI\otimes\frac{1}{2}\bbI.
\end{equation}

To consider the representation \eqref{seprep2},  we again use the fact that \eqref{pauli0}--\eqref{pauli3} imply that any initial single qubit $\rho(0)$  as time evolves tends to $\bbI/2.$ Therefore

\begin{equation}\label{asysep2}
\rho(t)=\sum_{k=1}^Kp_k\rho_{1,k}(t)\otimes\rho_{2,k}(t)\;\;\longrightarrow \rho(\infty) = (\sum_{k=1}^Kp_k)\frac{1}{2}\bbI\otimes\frac{1}{2}\bbI = \frac{1}{2}\bbI\otimes\frac{1}{2}\bbI.
\end{equation}

A side remark is the following. Suppose that the system is initially in a state like \eqref{seprep}, and that the density matrices correspond to pure states, but the second component is isolated from the thermal bath. Write the pure states of the second component as $\rho_{2,k}=\bbI+\bn_{2,k}\cdot\sigma,$ with $\|\bn_{2,k}\|=1.$ In equilibrium that state will be:

$$\rho(\infty) = \frac{1}{2}\bbI\otimes\big[\bbI+\big(\sum_k\mu_k\bn_{2,k}\big)\cdot\sigma\big].$$
That is, the thermal bath destroys the correlation between the two subsystems, but at the end, the isolated system may not be in a pure state unless $\|\sum_k\mu_k\bn_{2,k}\|=1.$

\subsection{The entropy change due to thermalization}
In the particular case of the first example, the entropy change due to thermalization is simple to compute. Using $S(\rho(t)) = -\tr\big(\rho(t)\ln\big(\rho(t)\big)\big),$ under the conventional $0\ln 0=0,$ we see from \eqref{ini1} and \eqref{ini1.2} that
$$S\big(\rho(0)\big) = \ln2,\;\;\;\;\mbox{whereas in the limit}\;\;\; S\big(\rho(\infty)\big) = \ln4 = 2\ln2.$$
Since a system in a thermal bath is anything except isolated, we can interpret the increase in entropy as related to the work done by the environment to decorrelate the system. 

\section{Final remarks}
The model developed above, is simple enough to allow for full analytical treatment. Its features contain the basic
ingredients that confirm the physical intuition of the situation, that is, if the environment affects the qubits independently, the system evolves to a trivial equilibrium state. 

A possible explanation for the fact that regardless of the initial sate of the qubit, in equilibrium the two levels are equally populated, lies in the nature of the effect of the environment. All that a random shock does, is to flip the microscopical state. So in the long run, its effect is to spread the initial probability evenly among the two levels.

The model applies as well to any number of non-interacting qubits, arbitrarily coupled through their initial state. Therefore, in a system of $N$ identical qubits, at equilibrium the total number of qubits in the upper level follows a binomial distribution $B(N,1/2).$

Notice that the representation \eqref{paurep} applies to any density matrix for the two qubit system, be it entangled or separable, thus under the Blume-Clauser model of thermal bath, the results in \eqref{asypau1} or \eqref{asysep2} provide a transparent way of understanding how a pair of entangled qubits become decorrelated by the action of the thermal bath.

\textbf{Conflict of interest statement} I hereby certify that the current work involves no conflict of interest of any kind. Also, as no data is used, there is no need to mention data sources.

\textbf{Data availability} There is no data availability issue associated to this submission.
%\textbf{Declaration of competing interests:} We have no competing interests to declare.

\end{document}